\documentstyle[12pt,aaspp4,flushrt]{article}

\lefthead{J. Sylvestre, J. Khoury et al.}
\righthead{Turbulence in YSOs: Multifractal Universality and GSI}

\begin{document}
\title{Turbulent Bipolar Outflows in Young Stellar Objects:\\ 
Multifractal Universality Classes and Generalized Scaling}
\author{Julien Sylvestre~\altaffilmark{1}, Justin Khoury~\altaffilmark{2} and Shaun Lovejoy}
\affil{Physics Department, McGill University, Montr\'eal, Qu\'ebec, Canada.}
\authoraddr{Julien Sylvestre, MIT NW17-161, 175 Albany St., Cambridge, MA 02139, USA.}
\altaffiltext{1}{Current address: Massachusetts Institute of Technology, Physics Department, 77 Massachusetts Avenue, Cambridge, MA 02139, USA; email address: julien@mit.edu}
\altaffiltext{2}{email address: jkhoury@princeton.edu}
\and
\author{Pierre Bastien~\altaffilmark{3}}
\affil{D\'epartement de Physique, Universit\'e de Montr\'eal, Montr\'eal, Qu\'ebec, Canada}
\altaffiltext{3}{Guest observer at the Canada-France-Hawaii telescope.}
\and
\author{Daniel Schertzer}
\affil{Laboratoire de Mod\'elisation en M\'ecanique, Universit\'e Pierre et Marie Curie, Paris, France}

\begin{abstract}
We present a study of the scaling properties of bipolar outflows, generic to the circumstellar environment of young stellar objects, and which are generally believed to be magnetohydrodynamically driven. Our work consists of a multifractal analysis of near-infrared light scattered by dust grains in the lobes of seven bipolar nebulae, namely GGD 18, V380 Orionis, LkH$\alpha$ 101/NGC 1579, LkH$\alpha$ 233, PV Cephei, V645 Cygni (GL 2789), and V633 Cassiopeiae (LkH$\alpha$ 198). We argue that these objects pertain to a multifractal universality class, with the corresponding universal parameters averaged over the ensemble yielding $\alpha=1.96\pm 0.02$, $C_1=0.04\pm0.02$, and $H=0.7\pm0.2$; 
these results might suggest that the dynamics of the observed objects are similar.
We then proceed to investigate the existence of anisotropy in the scaling of GGD 18 using the Generalized Scale Invariance (GSI) formalism. 
It is found that differential rotation and stratification are involved in the outflow mechanism.
The stratification is believed to be dynamical in nature i.e., gravity is apparently not the dominant stratifying force.
\end{abstract}


\section{Introduction} \label{intro}

The presence of bipolar outflows in the vicinity of young stellar objects (YSOs) is a generic feature of early stellar evolution. 
The outflows consist mostly of cold molecular gas, ionized material, and dust grains ejected with velocity in the range 100-300 km/s. A second characteristic of the circumstellar environment is a Keplerian accretion disk surrounding the protostar; the existence of these optically thick disks has been suggested by various observations, such as the high degree of linear polarization shown to be the result of multiple scattering by material in the disk (Bastien \& M\'enard 1990), and the absence of a red-shifted component in forbidden-line emission (Appenzeller et al. 1984; Edwards et al. 1987).

The observation that the thrust of the outflowing material cannot be radiatively driven (Bally \& Lada 1983; Cabrit 1989) suggests that the flows are hydromagnetic in nature. Moreover, experimental evidence of correlations between the physical properties of the disk and those of the ejected material hint that the accretion disk is a necessary ingredient in the driving mechanism. For example, continuum observations have indicated that YSOs with outflows of larger spatial extent correspond to stronger millimeter fluxes, which in turn can be related to the presence of a more massive disk (Cabrit \& Andr\'e 1991; Bontemps et al. 1996). Furthermore, there are correlations between the forbidden-line emission profiles which are generated in the outflows, and near-infrared excess believed to be related to the accretion flow evolution (Gomez de Castro \& Pudritz 1992; Edwards et al. 1987).

Current models for the origin of the outflows involve magnetohydrodynamics (MHD) winds driven by the accretion disk, and can be divided into two categories. The first class of models postulates that the winds are driven by a magnetic field threading through the disk (Pelletier \& Pudritz 1992; Ferreira \& Pelletier 1993a,b, 1995): particles rotating in the vicinity of the disk are subjected to a centrifugal force whose component perpendicular to the magnetic field lines may be neglected (assuming the magnetic field to be sufficiently strong); this results in a (centrifugal) acceleration along the field lines, thus generating the outflows. A second popular interpretation is the so-called X-celerator model (Shu et al. 1988, 1994a,b); in the region of the stellar equatorial plane where the radial acceleration is approximately null (due to counterbalance of the gravitational and centrifugal forces), the strong magnetic field of the YSO takes over the accretion mechanism, thus generating outflows. For the sake of briefness, we shall refer to these two categories of models as the PP and Shu models, respectively.

It is clear that further experimental data and analysis methods are
needed to clarify the situation. In this spirit, we believe that a study
of the scaling properties of the outflows could provide complementary
information on the dynamics involved. 
The search for scaling in the
outflows is motivated by the observation that over all scales, the equations of MHD for a non-dissipative medium present no characteristic length (e.g., Carbone 1993); in addition, since the law of gravity is a scale invariant (power law) form, the gravitational field of the disk-protostar system will not introduce a characteristic scale,
and hence {\it a priori} it is possible to obtain scale invariance over
all scales ranging from the macroscopic size of the lobes down to the
much smaller dissipation scale (i.e., the inner scale of the inertial range). 
A fundamental observation in the
analysis of three-dimensional MHD turbulence is the existence of three physical
quantities, namely the total energy (i.e. the sum of the magnetic and the kinetic energies), the cross-helicity, and the
magnetic helicity~\footnote{The cross-helicity and magnetic helicity
densities are defined by $\vec{v}\cdot\vec{B}$ and
$\vec{A}\cdot\vec{B}$, respectively, where $\vec{v}$ is the velocity
field, $\vec{B}$ is the magnetic field, and $\vec{A}$ is the vector
potential.}, which are conserved by the non-linear terms in the ideal
equations (i.e., without forcing terms, nor dissipation). 
Considering the
disk-protostar system as injecting the corresponding fluxes at large scales, they subsequently propagate (or cascade) towards smaller scales until the scale where dissipation becomes important is reached. 
Furthermore, such propagation mechanisms are more efficient
between scales of similar magnitude (usual MHD turbulence is
correspondingly ``local'' in Fourier space; e.g., Biskamp 1993; Carbone
et al. 1996), and over the inertial range, the (statistical) laws governing the propagation of flux are scale invariant. The combination of conserved fluxes and local interactions (in scale) is the basis for phenomenological cascade models of turbulence. 
In the case of YSOs, since MHD turbulence implies the existence of three conserved fluxes in the inertial range, the dynamics of the outflows can approximatively be described by three non-linearly coupled cascade processes (see Schertzer \& Lovejoy 1995; Schertzer et al. 1997b), and, in the approximation that dust grains constitute a passive scalar quantity (i.e., they are advected by the velocity field without disturbing it), there will be an additional coupled cascade corresponding to the conserved flux of passive scalar variance (Obukhov 1949; Corrsin 1951). 

While the idea of cascades in hydrodynamic turbulence was introduced by
Richardson (1922), the first explicit cascade models were not developed until
the 1960's (Novikov \& Stewart 1964; Yaglom 1966; Mandelbrot 1974), and
have become since then the basic tools for studying turbulent
intermittency. Developments in the following two decades led to the
current understanding that cascade models generically produce
multifractals (Schertzer \& Lovejoy 1985, 1987b), hence establishing their relevance in analyzing and modeling scale invariant multifractal fields. Indeed, multifractals have already been applied to various astrophysical problems such as the large-scale structure of the universe (e.g., Wiedenmann et al. 1990; Coleman \& Pietronero 1992; Borgani et al. 1993; Martinez \& Coles 1994; Garrido et al. 1996; Sylos Labini \& Montuori 1998; Sylos Labini et al. 1998; Lovejoy, Garrido \& Schertzer 1999), Ly$\alpha$ clouds (Carbone \& Savaglio 1996), the cosmic microwave background radiation (Pompilio et al. 1995), photospheric magnetic fields (Cadavid et al. 1994), photometric data of NGC 4151 (Longo et al. 1996), and the solar wind (Carbone 1993, Politano \& Pouquet 1995). 

A basic difficulty in multifractal analysis and modeling is that at a
purely general level, multifractals implicitly involve an infinite
number of parameters (e.g., the codimension function), and hence would
be unmanageable if no further simplification could be made. Fortunately,
there exist multifractal universality classes, which are stable
attractors of multiplicative cascades (Schertzer \& Lovejoy 1987b,
1989a,b, 1991, 1997a). Universality is of practical importance, since it reduces the number of parameters required for the description of the scaling function of multifractal fields to only three.

Although the cascades and corresponding multifractals usually discussed in the literature involve isotropic scaling, physical systems generally exhibit preferred spatial directions (for example, almost all scaling geophysical systems are strongly stratified due to gravity). The need to handle scaling freed from the constraints of isotropy led to the development of the formalism of Generalized Scale Invariance (hereafter GSI, Schertzer \& Lovejoy 1985, 1987a,b, 1989a,b, 1991; Lovejoy \& Schertzer 1985) which is the most general framework for describing anisotropic scaling. In the case of bipolar nebulae, the application of GSI to the study of their scaling properties is motivated by the observation that the outflowing material is not isotropically ejected from the disk-protostar system, and the anticipation that the direction and strength of the anisotropy may vary with scale.

In this paper, we present a multifractal analysis of near-infrared light
scattered from dust grains in bipolar outflows, using images of V380
Orionis, V645 Cygni (GL 2789), LkH$\alpha$ 101/NGC 1579, LkH$\alpha$
233, PV Cephei (for a discussion of their physical parameters, see Bastien \& M\'enard 1990), V633 Cassiopeiae (e.g., Asselin et al. 1996), and GGD 18 (Gyulbudaghian et al. 1978). 
An immediate issue in interpreting scattered light field is the
relationship of radiative transfer to the density field of the
lobes; indeed, this basic remote sensing problem of radiative transfer
through multifractal clouds constitutes an important application of
multifractals (e.g., Lovejoy et al. 1995; Lovejoy \& Schertzer
1995). 
Since the
radiative transfer equation has no characteristic scale, if the spatial
distribution of scatterers is scale invariant so will the related
radiation field. It is therefore reasonable to infer  that if the scattered light field displays scale invariance over a wide range of scales, then the underlying matter distribution will also possess scale invariant statistics. 
We shall come back to previous attempts of quantifying the correlations between the respective multifractal parameters of the scattered light and matter fields.

Information concerning the acquisition of CCD images is provided in section 2. In section 3, the basic concepts of universality are presented, while section 4 summarizes the results of the universal parameters for the seven objects in the ensemble. An overview of the GSI formalism is the subject of section 5, while the measured anisotropy parameters of GGD 18 are presented in section 6. Finally, we discuss in section 7 the potential implications of similar analyses on current and future models of star formation.

\section{Observations} \label{data}

Images of V380 Orionis, V645 Cyg (GL 2789), LkH$\alpha$ 101/NGC 1579,
LkH$\alpha$ 233, PV Cep, and V633 Cas were obtained at the f/15 focus of
the 1.6m Ritchey-Chr\'etien telescope at the {\it Observatoire Astronomique du Mont M\'egantic} (hereafter OMM). The first five objects were observed on 1997 October 23 in the I bandpass using a 2048x2048 pixel$^2$, 16-bit Loral CCD camera with a scale of 0.13'' per pixel;
with this optical configuration and at the time of observation, the seeing was of the order of 1.5''. On the other hand, V633 Cas was observed on 1989 September 25 in the I bandpass using a 512x320 pixel$^2$ RCA chip with 0.48'' per pixel. 
The seeing for these images was of the order of 1.3''.
Images of GGD 18 were obtained at the f/8 focus of the 3.6m
Canada-France-Hawai Telescope (hereafter CFHT) with the RCA2 1024x640 pixel$^2$ CCD for a scale of
0.108'' per pixel, during the period 1987 December 23-28;
the seeing for these images was 0.5''.
All images were processed to correct for cosmetic defects of the CCDs,
for reading noise, and for cosmic rays.
The resolution of images of GGD 18 was increased by deconvoluting the
images with the method of maximal entropy (Gravel 1990), resulting in an effective seeing of 0.39''. Acquired images and
exposures for all the afore-mentioned objects are summarized in Table~\ref{tab:list_objects}.
\begin{deluxetable}{lccc}
\tablecaption{Observations \label{tab:list_objects}}
\tablehead{\colhead{Object} & \colhead{R.A. (2000)} & \colhead{Dec. (2000)} & \colhead{exposure (min.)}}
\startdata
GGD 18	& 06:34:35.72	& +04:12:43.8	& 120	\nl
V380 Ori	& 05:36:25.31	& -06:42:57.11	& 30	\nl
LkH$_\alpha$ 101/NGC 1579 & 04:30:14.34	& +35:16:24.1	& 45	\nl
LkH$_\alpha$ 233 & 22:34:40.85	& +40:40:04.61	& 80 \nl
PV Cep	& 20:45:54.00	& +67:57:35.80	& 60 \nl
V645 Cyg (GL 2789)       & 21:39:58.16    & +50:14:21.5     & 60	\nl
V633 Cas	& 00:11:26.43	& +58:49:50.04	& 7	\nl
\enddata
\end{deluxetable}

\begin{figure}
\begin{center}
\plotone{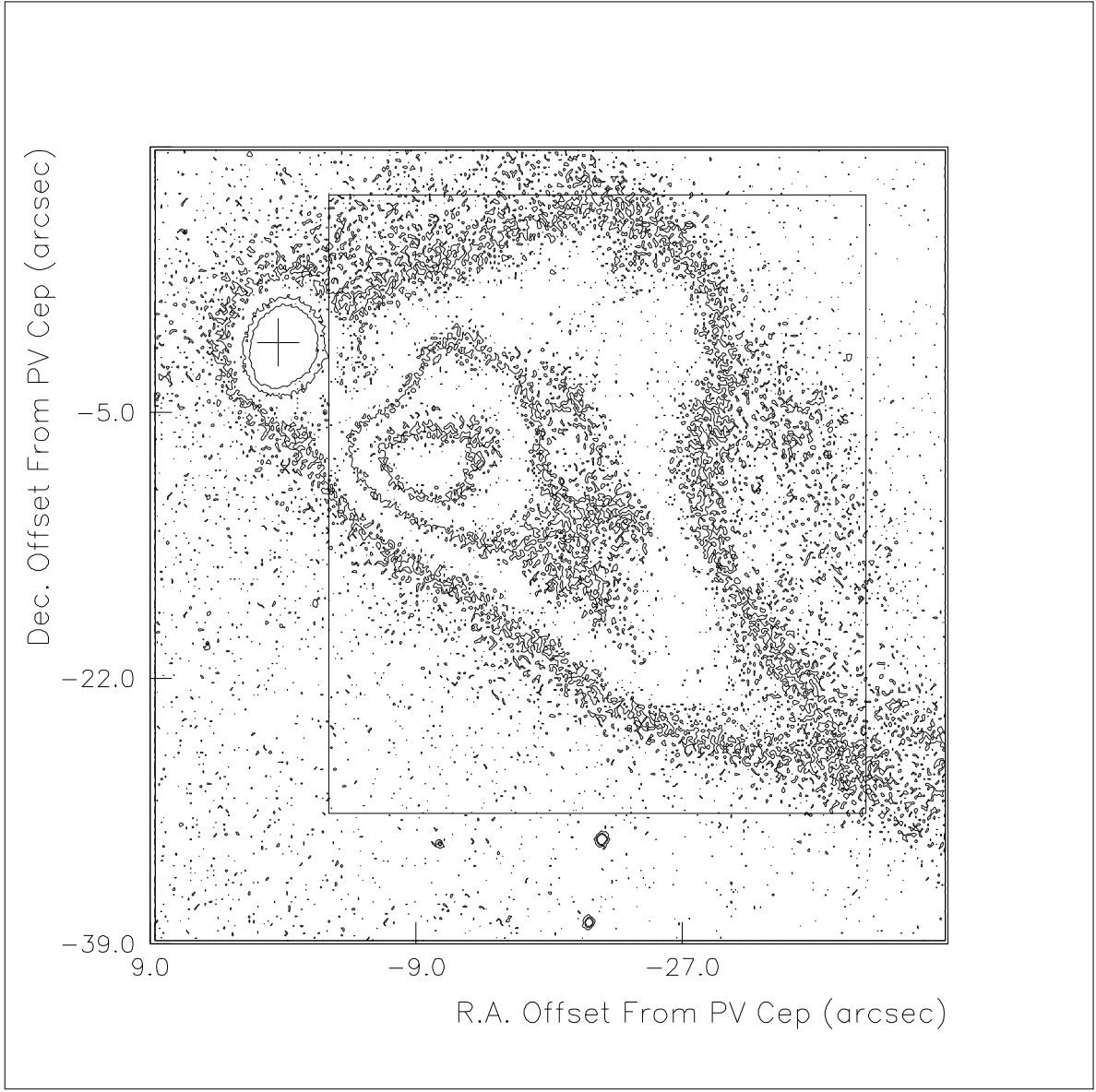}
\end{center}
\figcaption[f1.eps]{Contour plot of PV Cep observed at OMM in the I filter, where contours are scaled linearly down to $2\sigma$. The approximate location of the protostar is indicated by the cross, while the superimposed box delimits the region selected for the analysis. North is up, and east is to the left. 
\label{fig:pvcep_subsection}}
\end{figure}

As YSOs generically occupied only a fraction of the region of observation, it was necessary to cut the images into sections for the purpose of our analysis. Being physically distinct from the dust shells, background stars as well as the protostar were always excluded from these sub-images since their presence would bias the analysis of scattered light. 
Sections were also chosen to be far enough from the protostar to avoid problems of large intensity gradients associated with the protostar, such that the cloud--radiation physics (and hence statistics) for a given sub-image could be considered approximately invariant under translations.
Finally, regions where the intensity was not significantly greater than the background noise were also excluded.
A contour plot of a sample CCD image acquired at the OMM, namely PV Cep, is shown in Figure~\ref{fig:pvcep_subsection}, where the box delimits the section used for the analysis.

\begin{figure}
\begin{center}
\plotone{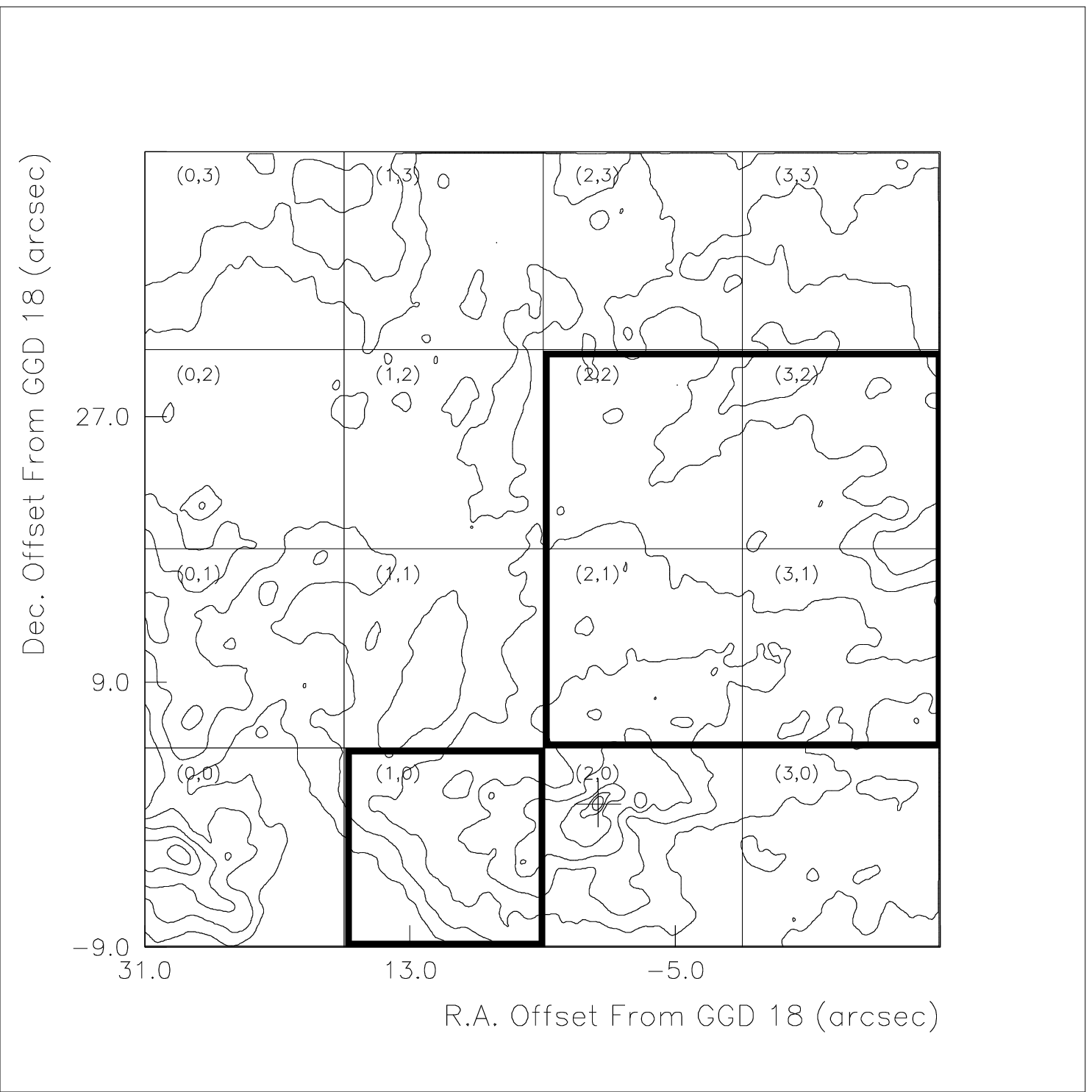}
\end{center}
\figcaption[f2.eps]{Contour plot of GGD 18 observed at the CFHT in the I filter. Contours correspond to intensities of $\sigma/81$, $\sigma/3$, $\sigma$, $3\sigma$, $9\sigma$ and $27\sigma$. The approximate location of the source is identified by the cross. Superimposed is a grid defining sub-images (and corresponding labels) used for the subsequent analysis. The purpose of the darkened edges will be explained in section~\ref{sig}. North is up, and east is to the left.
\label{fig:ggd18_image}}
\end{figure}

Figure~\ref{fig:ggd18_image} presents a contour plot of the region of GGD 18 analyzed, divided into sub-regions that will be used for the subsequent multifractal analysis (sections~\ref{dtm} and~\ref{sig}).
The image of GGD 18 was the only one in the ensemble with sufficient intensity and spatial resolutions to allow a GSI analysis (see section~\ref{sig}), and it is thus worthwhile to describe some of its known features. 
This object is located at 30.25'' NW of the binary GL 961 (Cohen 1973), and was first identified as a possible candidate for a HH object (Gyulbudaghian et al. 1978). 
A few years later, Lenzen et al. (1984) discovered an IR source embedded in the nebulous region of GGD 18 (it corresponds to the maximum of intensity of Figure~\ref{fig:ggd18_image}, identified by a small cross), but did not speculate on its nature. 
Polarization maps from deconvoluted images revealed (Gravel 1990) a centrosymmetric pattern centered on the GGD 18 source showing that it was a YSO, along with a region of aligned vectors near the source indicating the presence of an optically thick disk (the annular structure near the source and along the NE and SW directions is believed to correspond to a fraction of the disk that is visible); the inclination of the latter with respect to the line of sight was estimated at $30^o$ (the northern lobe points toward the observer), and the symmetry axis of the nebula points at $30^o$ NW (Gravel 1990). Finally, spiral-like polarization patterns were discovered in the vicinity of the source, indicating a complex distribution of matter that may be the result of rotation in the material near the source, combined with the effects of possible local inhomogeneities in the disk. 
An unresolved issue is the extent at which the dynamics of the outflows are influenced by the proximity of the binary GL 961; evidence that the CO bipolar jet of GL 961 reaches and influences the outflow dynamics of GGD 18 was provided by Gravel (1990), but a quantitative study of such interactions is still needed.
To summarize his argument, we notice that both lobes are more extended in the western direction, and the apparent contraction on the eastern side is believed to result from interactions with the jet; the latter appears as a faint island (approximately 12'' in length) located in (1,1), and points in the NW direction. It is more visible in (1,2) and enters both (1,3) and (2,3), where it turns around towards the SW direction; such deviations could be the result of interactions with nebulous structures to the north of GGD 18. 
The local maximum in section (0,0) corresponds to the western component of the binary GL 961; since GL 961 is an intense source of unscattered light, the region (0,0) will be excluded to allow an unbiased analysis of scattered light in the outflows. 
For similar reasons, section (2,0) will also be excluded from the analysis.

\section{Review of Multifractal Processes} \label{dtmtheory}

\subsection{Universal Multifractals}
It is well known that the Navier-Stokes (NS) equations are invariant under the rescaling  $x \rightarrow x \lambda^{-1}$, $v \rightarrow v \lambda^{-H}$, and $t \rightarrow t\lambda^{H-1}$. 
Assuming that $\epsilon$, the energy flux to smaller scales, is a scale invariant quantity, it is found that $H=1/3$, and dimensional analysis leads to the famous scaling law $E(k)\propto k^{-5/3}$ for the energy density in momentum space (Komolgorov 1941; hereafter K41). 
While the equations of MHD satisfy similar scaling relations, there is no consensus on what the analogue of K41 should be (the corresponding dimensional analysis no longer gives a unique dimensional combination, hence unique exponent). 
In terms of Elsasser variables, eddies fall into two classes depending on their direction of propagation along the magnetic field lines: interactions between eddies belonging to different classes are less likely, thus weakening the energy transfer (Iroshnikov 1963; Kraichman 1965; hereafter IK). 
Consequently, the characteristic interaction time $\tau_{eddy}$ (i.e. the eddy turnover time) is increased to $(\tau_{eddy}/\tau_A)^a$, where $\tau_A$ is the characteristic time for Alfv\'en waves, and $a$ is some positive constant (Politano \& Pouquet 1995). 
The scaling relation becomes:
\begin{equation}
E(k) \propto k^{-\left(1+\frac{2}{a+3}\right)},
\end{equation}
\noindent with $a=1$ corresponding to the IK theory.

It is found experimentally that in hydrodynamical media K41 is generally not respected for individual realizations -- even in estimates of the ensemble average, the exponent differs from 5/3; the discrepancy can be attributed to intermittency (fluctuations in $\epsilon$ due to small-scale non-linear structures). 
As discussed in the introduction, multiplicative cascades model the propagation of conserved fluxes in turbulent intermittent media, and in general present the expected characteristics of a fully developped turbulent field.
The general outcome of such cascades is a multifractal field which is scale invariant over the inertial range, and whose flux density at resolution $\lambda$, denoted by $\epsilon_\lambda$, is described by (Schertzer \& Lovejoy 1987b):

\begin{equation}
\langle \epsilon_\lambda^q\rangle = \lambda^{K(q)},  \label{eq:kq}
\end{equation}

\noindent where the brackets indicate an average over many realizations, and $K(q)$ is the moment scaling function. 
The arbitrariness of $H$ in the rescaling of the NS or MHD equations allows the possibility of scaling of different moments of the intensity spectrum, and this feature is what equation~\ref{eq:kq} describes.

While in general $K(q)$ need only be convex, cascades possess stable, attractive {\it universality classes} (Schertzer \& Lovejoy 1987b, 1997) whose description requires only three parameters, namely $\alpha$, $C_1$ and $H$, with the corresponding moment scaling function determined by the universality relation:
\begin{equation}
K(q)-qH=\frac{C_1}{\alpha-1}(q^\alpha-q).
\label{eq:mf_universality}
\end{equation}
The significance of each of the three universality parameters on the multifractal field can be described as follows:
\begin{itemize}
\item $C_1$ corresponds to the codimension of the mean field, and thus
distinguishes between a field whose mean is dominated by a few localized
intense peaks (large $C_1$), and one with a mean dominated by a larger
proportion of its surface (small $C_1$ --- for non-fractal such as white noise, $C_1=0$);

\item $H$ is a measure of the degree of (scale by scale) non-conservation of the field, or qualitatively a measure of its smoothness (with large values of $H$ corresponding to smoother fields, see eqs.~\ref{eq:iso_scaling} and~\ref{eq:H}). For example, in usual hydrodynamic turbulence the energy flux to smaller scales is conserved ($H$=0) whereas the velocity shears have the Komolgorov value $H=1/3$;

\item $\alpha$ is the degree of multifractality, i.e., a measure of the
deviation from the monofractal case. As $\alpha$ is the L\'evy index of
the multifractal generator, we have the restriction $0 \leq \alpha \leq
2$, with $\alpha=0$ and $\alpha=2$ corresponding to monofractal ($\beta$
model) and log-normal models, respectively~\footnote{Note that
the frequently used expressions ``log-L\'evy'' and ``log-normal'' are rather misleading because of the divergence of high order statistical moments;
the statistics will only be approximately ``log-normal'' and ``log-L\'evy'' up to the given critical order of divergence of the moments.}.
\end{itemize}

\subsection{Analysis of Physical Fields}
A preliminary verification of the existence of scale invariance is that the spectral energy density satisfies a general (isotropic) scaling law:  
\begin{equation}
E(k) \propto k^{-\beta},
\label{eq:iso_scaling}
\end{equation}
\noindent where $\beta$ is the scaling exponent, or spectral slope.

After the existence of a scaling regime is established, one can proceed to compute the scaling function $K(q)$ and test for universal multifractal behavior.
An efficient technique for that purpose is the Double Trace Moment (DTM) method (Lavall\'ee 1991; Lavall\'ee et al. 1991, 1993).
A new function $K(q,\eta)$ is first defined similarly to $K(q)$ in equation~\ref{eq:kq}, but for a field $\epsilon_\Lambda$ at its maximal resolution $\Lambda$, raised to the $\eta$ power (i.e., $\epsilon_\Lambda \rightarrow \epsilon_\Lambda^\eta$), and renormalized by spatial averaging. Hence, writing $\langle (\epsilon_{\Lambda}^{\eta})_{\lambda}^q \rangle$ to indicate the $\lambda$-resolution $q^{\rm{th}}$ moment of $\epsilon_{\Lambda}^{\eta}$, we obtain the following generalization of equation~\ref{eq:kq}:

\begin{equation}
\langle (\epsilon_{\Lambda}^{\eta})_{\lambda}^q
\rangle=\lambda^{K(q,\eta)}. \label{eq:newflux}
\end{equation}

\noindent While it can be shown (Lavall\'ee 1991) that $K(q,\eta)$ and $K(q)$ are related by
\begin{equation}
K(q,\eta)=K(q\eta)-qK(\eta), \label{eq:Kqetab}
\end{equation}
the advantage of the DTM technique for testing and characterizing
universality compared with other methods (e.g., Schmitt et al. 1995) is realized when $K(q)$ is universal (eq.~\ref{eq:mf_universality}), in which case the $\eta$-dependence factorizes:
\begin{equation}
K(q,\eta)=\eta^\alpha K(q). \label{eq:Kqeta}
\end{equation}

\noindent The DTM method allows the computation of $K(q,\eta)$, and assuming universality, the L\'evy index $\alpha$ can be deduced from equation~\ref{eq:Kqeta}. The remaining parameters $C_1$ and $H$ are also determined from a knowledge of $K(q,\eta)$. Explicitly, one finds:

\begin{equation}
C_1=(\alpha -1) \cdot \frac{K(q,1)}{q^{\alpha}-q}, \label{eq:c1}
\end{equation}
\noindent and

\begin{equation}
H=\frac{\beta-1}{2}-C_1\cdot\frac{2^{\alpha-1}-1}{\alpha-1}.
\label{eq:H}
\end{equation}

We conclude this section with a few comments concerning the range of
validity of the above equations when analyzing images of physical
fields. 
In practice, the finite size of the sample implies that sufficiently high order moments are dominated by the largest value assumed by the field, and therefore underestimate the true ensemble moments. 
Beyond some threshold $q_s$, which for a single realization is given by
\begin{equation}
q_s=\left(\frac{d}{C_1}\right)^{1/\alpha}, \label{eq:q_s}
\end{equation}
equation ~\ref{eq:mf_universality} is no longer expected to hold true, and $K(q)$ becomes linear.
Here, $d$ is the dimension of the space over which the analysis is made ($d=2$ for the images discussed here).
One therefore encounters a ``multifractal phase transition'' (Schertzer et al. 1992).

\section{Double Trace Moments Results} \label{dtm}

\begin{figure}
\begin{center}
\plotone{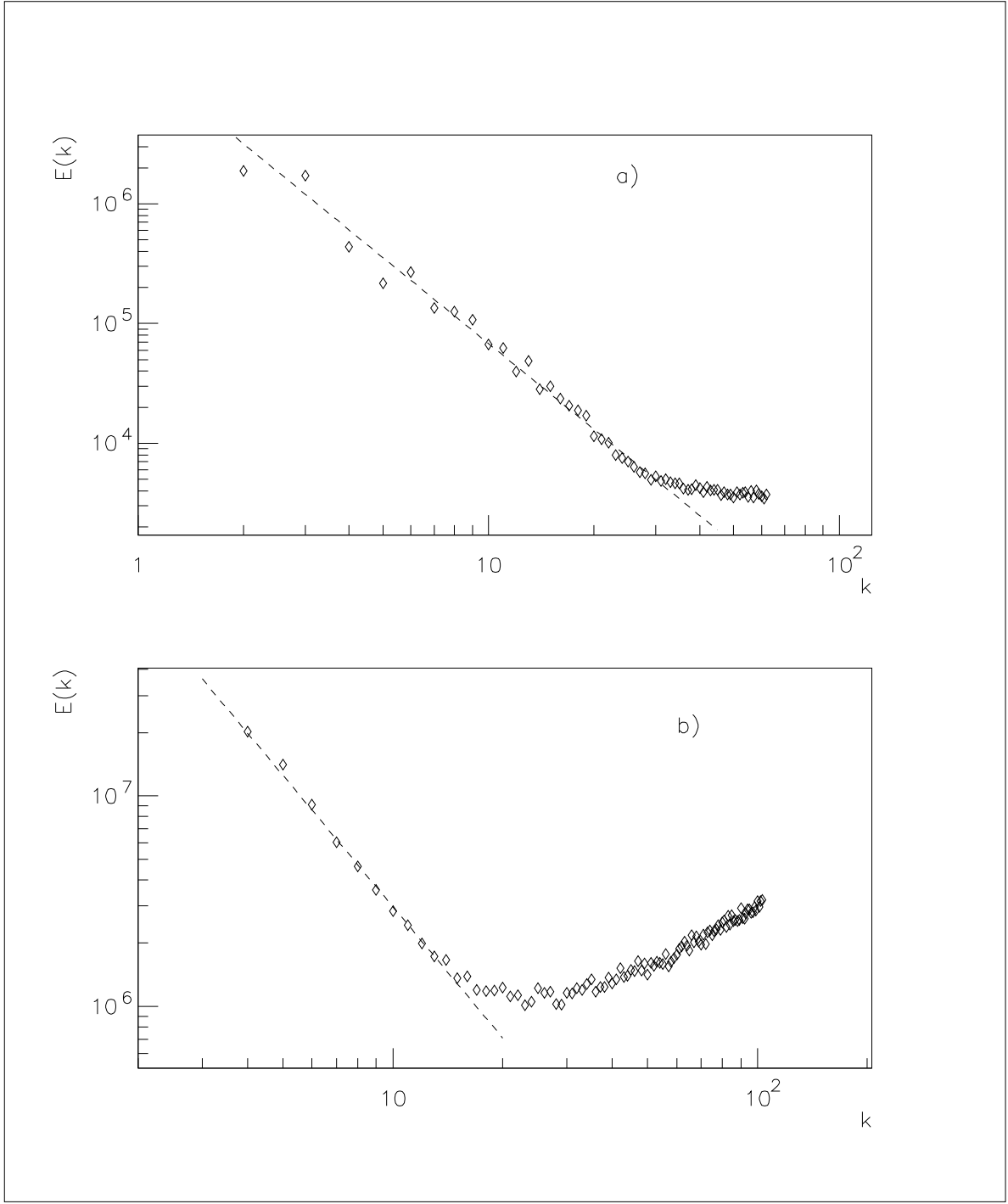}
\end{center}
\figcaption[f3.eps]{a) Power spectrum of sub-region (1,3) of GGD 18
(fig.~\ref{fig:ggd18_image}) yielding $\beta$=2.4$\pm$0.3; b) Power spectrum of a sub-region of
LkH$\alpha$ 101 with $\beta$=2.0$\pm$0.2; the high-frequency region
corresponds to $\beta\approx -1$, as expected for Gaussian white noise
in 2D space. 
\label{fig:powerspec}}
\end{figure}

As discussed in section~\ref{dtmtheory}, scale invariance is a necessary
condition for a physical field to be multifractal, and its existence
should be verified before computing the universal parameters. A first
analysis is to consider the isotropic power spectrum since the
spectral energy density $E(k)$ of an isotropic MHD turbulent medium is
expected to obey a scaling law with exponent $\beta$ (see
eq.~\ref{eq:iso_scaling}). Examples of the power spectra obtained are
shown in Figure~\ref{fig:powerspec}a) and b) for a sub-region of GGD 18
and LkH$\alpha$ 101/NGC 1579, respectively, where $E(k)$ has arbitrary
units. 
In each case, we note a break in scaling occurring at small
scales (i.e., large wavenumber), followed by a regime of constant or
increasing energy density. We shall argue that the scale at which
scaling breaks corresponds to the resolution at which structures of the
intensity field become dominated by noise. This is easily seen in the
case of GGD 18, where the break occurs at a resolution of approximately
0.3'', while the seeing at the time of data acquisition was estimated at
0.4''. On the other hand, the high frequency behavior of $E(k)$ for the
spectrum of LkH$\alpha$ 101 is linear ($\beta\approx -1$) in wavenumber, as expected for Gaussian white noise in 2D space~\footnote{In the case of LkH$\alpha$ 101, the fact that the break is not a manifestation of the seeing results from the limited exposures compared to those of GGD 18.}. These two sample power spectra illustrate a feature common to all power spectra in the ensemble, namely isotropic scaling down to the scale where noise dominates the statistics (the scaling anisotropy -- see sections \ref{gsitheo} and \ref{sig} -- is removed by the angular integration). 
Least-squares fits over the respective linear regions of Figure~\ref{fig:powerspec}a) and b) yield $\beta$=2.4$\pm$0.3 and 2.0$\pm$0.2, respectively, where the uncertainties are estimated from the fitting procedure.

\begin{figure}
\begin{center}
\plotone{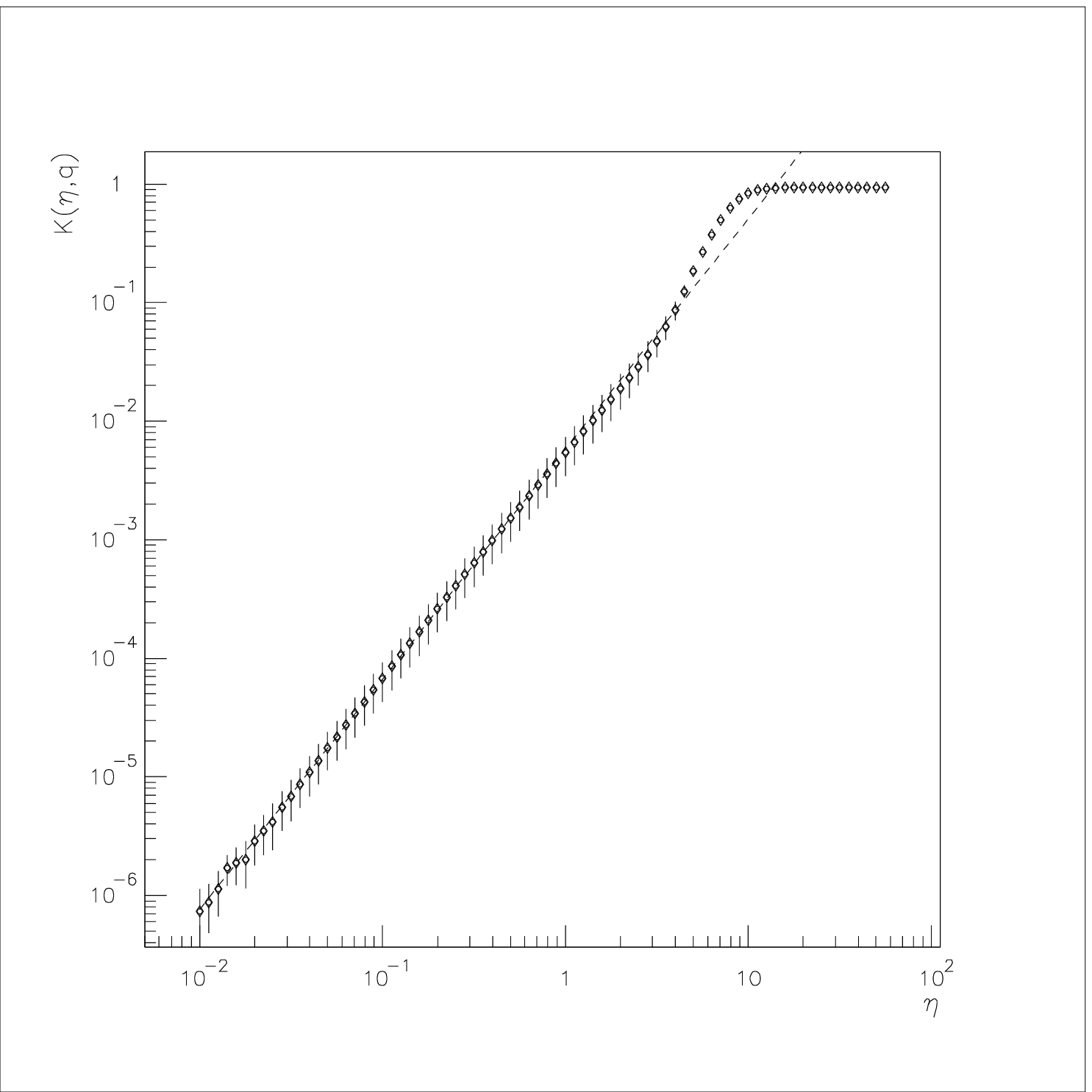}
\end{center}
\figcaption[f4.eps]{Plot of $K(\eta,q)$ obtained with a DTM analysis of a sub-region of LkH$\alpha$ 233 using
$q=0.5$, where the error bars correspond to uncertainties in the least-squares fits performed to obtain each point. 
From the slope and the $\log\eta=0$ intercept of the linear region, we obtain $\alpha = 1.93$, and $C_1 = 0.023$. There is a first order multifractal phase transition near $\eta=4$, which is a consequence of the finiteness of the analyzed sample.
\label{fig:trace_dtm}}
\end{figure}

With the existence of scaling established, a DTM analysis was performed
on each sub-image in the ensemble using four values of $q$, namely
$q$=0.5, 0.6, 0.75, and 0.9. As an example of the DTM results, Figure~\ref{fig:trace_dtm} presents $K(\eta,q)$ with $q=0.5$ for a sub-image of LkH$\alpha$ 233, from which we note a power-law dependence over the range 0.06$\leq\eta\leq$1, in agreement with equation~\ref{eq:Kqeta}. 
The slope of the linear region yields $\alpha = 1.93$, while the intercept $K(q,1)$ gives $C_1 = 0.023$ (eq.~\ref{eq:c1}). Finally, there is a departure from the power-law dependence in the range $4\leq\eta\leq 10$, beyond which $K(\eta,q)$ becomes independent of $\eta$. As explained in section~\ref{dtmtheory}, equation~\ref{eq:Kqeta} breaks down beyond max$(q\eta,\eta)\approx q_s$ due to the finite size of analyzed samples. To confirm that this is indeed the cause for the multifractal phase transition, the substitution of the above values of $\alpha$ and $C_1$ in equation~\ref{eq:q_s} gives $q_s\sim 10$; on the other hand, $K(\eta,q)$ becomes horizontal at $\eta\approx10$, which implies that max$(q\eta,\eta)=10$ (recall that $q=0.5$), as expected.

Variations of the universality parameters over different sub-regions of a given nebula were in general observed to be within the known uncertainties, and this was explicitly verified in the case of GGD 18. Consequently, it was reasonable to average the universal parameters over all sub-regions for each object, and the results are summarized in Table~\ref{tab:results.all}. 
\begin{deluxetable}{lccc}
\tablecaption{DTM Results \label{tab:results.all}}
\tablehead{\colhead{Object} & \colhead{$\alpha$}	& \colhead{$C_1$}	& \colhead{$H$}}
\startdata
GGD 18	& $1.98$	& $0.075$	& $0.98$ \nl
V380 Ori & $1.97$	& $0.030$	& $0.51$ \nl
LkH$\alpha$ 101/NGC 1579  & $1.97$ & $0.031$	& $0.53$ \nl
LkH$\alpha$ 233	& $1.93$	& $0.023$	& $0.56$ \nl
PV Cep	& $1.92$	& $0.027$	& $0.52$ \nl
V645 Cyg (GL 2789) & $1.95$	& $0.036$ & $0.69$ \nl
V633 Cas & $1.97$	& $0.045$ & $0.89$ \nl
ENSEMBLE AVG.         & $1.96\pm0.02$	& $0.04\pm0.02$	& $0.7\pm0.2$ \nl
\enddata
\end{deluxetable}
Note that the accuracy on the parameters of individual objects can be roughly estimated from the accuracy of the DTM method (measured from numerical simulations in Lavall\'ee et al. 1991) to be $\pm0.1$ for $\alpha$, $\pm0.05$ for $C_1$, and $\pm 0.2$ for $H$. 
An immediate observation from Table~\ref{tab:results.all} is that $\alpha$ is very close to 2 for every object in the ensemble, a value which corresponds to the highest degree of multifractality. A second observation is the uniformity of the parameters over the ensemble, allowing one to compute ensemble averages listed in Table~\ref{tab:results.all}, where the uncertainties quoted for the ensemble averages correspond to one standard deviation from the mean. While these parameters describe the statistics of the field of scattered light, it is important to consider their relationship with those of the underlying field of matter; simulations (Naud et al. 1996) and simple theoretical arguments (Schertzer et al. 1997) have suggested that only $H$ is significantly affected by the scattering process, with the radiative value observed to be larger (corresponding to a smoother texture).

\begin{figure}
\begin{center}
\plotone{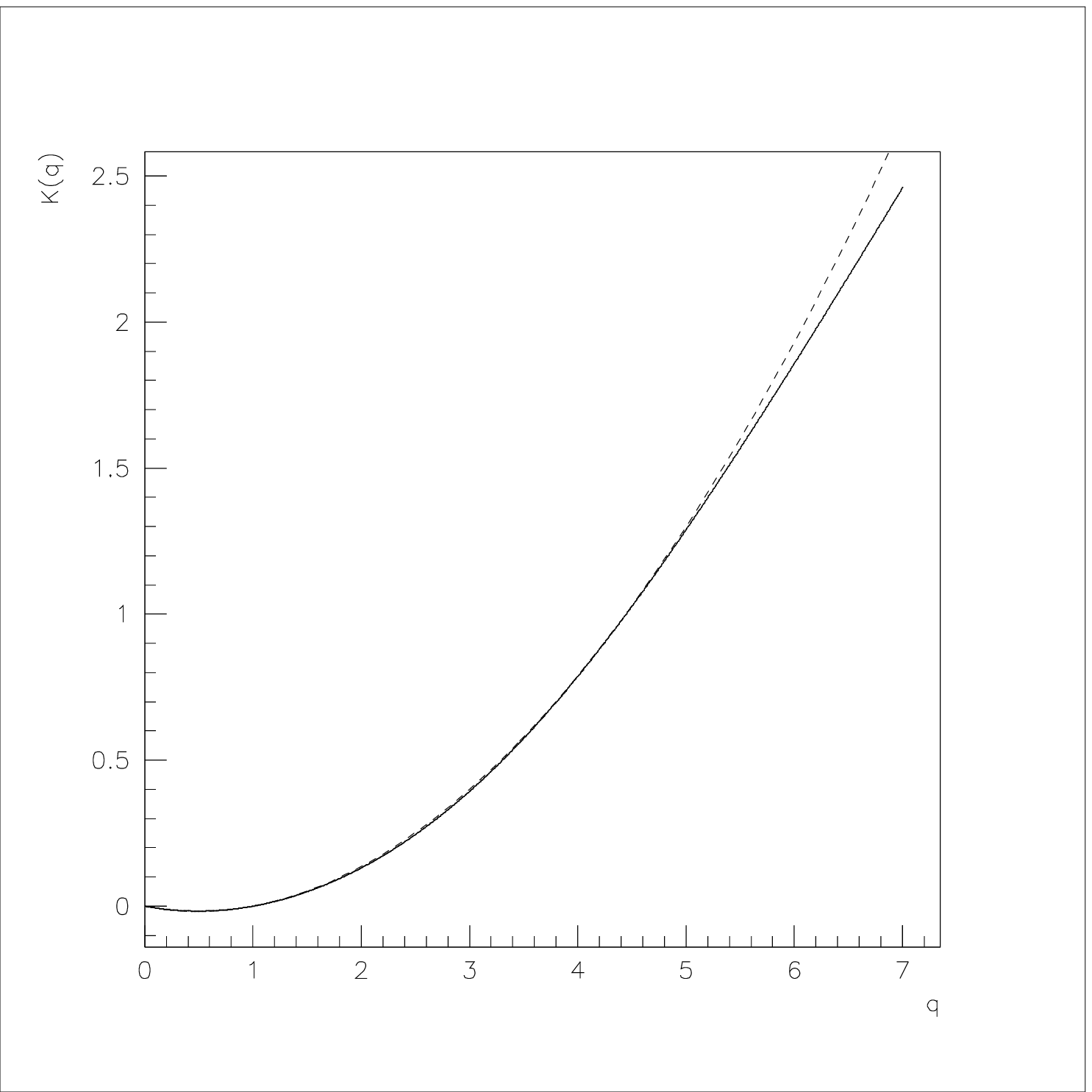}
\end{center}
\figcaption[f5.eps]{A comparison of a direct calculation of $K(q)$ (dashed line)
with the prediction of the universality relation
(eq.~\ref{eq:mf_universality}) for a sub-region of GGD 18. The universal
parameters used for the theoretical curve are $\alpha=1.93$ and
$C_1=0.045$, as calculated using the DTM method for the given sub-region. The linear behavior for $q\geq 5$, corresponding to a (first order) multifractal phase transition, is a result of the finiteness of the sample.
\label{fig:ggd.kq}}
\end{figure}

Recall from section~\ref{dtmtheory} that equation~\ref{eq:Kqeta} follows from equation~\ref{eq:Kqetab} provided that the scaling function $K(q)$ is universal. To test the validity of this assumption, one can compute $K(q)$ directly and compare the result with the prediction of the universality relation (eq.~\ref{eq:mf_universality}). Such a comparison is illustrated in Figure~\ref{fig:ggd.kq} for a sub-region of GGD 18. 
According to Figure~\ref{fig:ggd.kq}, we note that the two curves agree up to $q\approx5$, beyond which $K(q)$ becomes linear; from equation~\ref{eq:q_s} we find $q_s\approx 6$, indicating that the linear dependence is another manifestation of the finiteness of the sample. Similar agreement was observed for the other images in the ensemble.

\section{Generalized Scale Invariance} \label{gsitheo}

Our discussion has so far been constrained to the case of isotropic, or self-similar, scale invariance, for which structures at different scales are related by isotropic magnifications. However, more general anisotropic scaling ``zooms'' are possible and may arise as a result of forces inducing stratification and differential rotation in the dynamics, for instance. Rather than imposing, {\it a priori}, an isotropic notion of scale, the latter may be determined by the nonlinear dynamics; in this section, we attempt to empirically characterize this anisotropic scaling.

The need for a more general framework led to the Generalized Scale
Invariance (GSI) formalism, in which three ingredients are necessary for the description of scaling: 
\noindent (i) a unit ball $B_1$ consisting of all vectors of unit length, which can implicitly be defined by:
\begin{equation}
B_1 \equiv \{\vec{x} : g_1(\vec{x})\leq 1\}, \label{eq:unitball}
\end{equation}
\noindent where $g_1$ determines the notion of unit length; 
\noindent (ii) a scale changing operator $T_{\lambda}$ mapping a vector between two scales of ratio $\lambda$; once a unit ball is specified, all other scales can be identified by repeated applications of $T_{\lambda}$, thus generating a family of balls, $B_{\lambda}$; 
\noindent (iii) a definition of a measure of the $B_{\lambda}$, such
as the volume or a power of the volume.

It follows from the definition of $T_{\lambda}$ that these operators form a one-parameter multiplicative group, with corresponding generator $G$ given by:

\begin{equation}
T_{\lambda}=\lambda^{-G}. \label{eq:tlambda}
\end{equation}

\noindent While equation~\ref{eq:tlambda} allows $G$ to be non-linear
(Schertzer \& Lovejoy 1985), in order to simplify the
computations we shall make a linear approximation by assuming that $G$
and $T_{\lambda}$ are real matrices with constant coefficients, which is equivalent to assuming that the anisotropy of the scales is translationally invariant within the image analyzed. 
Since GSI analyses are
conventionally performed in Fourier space, the operator of interest is
the Fourier analogue of $G$, which shall also be denoted $G$ by an abuse
of notation (in the case of linear GSI, these matrices are transpose of
each other). Expanding the matrix generator in terms of pseudo-quaternions, we write:

\begin{equation}
G=1+f\sigma_x-ie\sigma_y+c\sigma_z=\left( \matrix{ 1+c & f-e \cr f+e & 1-c} \right), \label{eq:Gexpansion}
\end{equation}

\noindent where $1$ denotes the unit matrix, $\sigma_x$, $\sigma_y$, and $\sigma_z$ are the $SU(2)$ generators in the Pauli representation, and $c, e ,f$ are real (note that since the coefficient of the identity matrix would correspond to an isotropic magnification, it has been set equal to unity without loss of generality). This choice of basis is particularly convenient for GSI analysis as it decouples rotation and stratification. It follows from equations~\ref{eq:tlambda} and~\ref{eq:Gexpansion} that, within the approximation of linear GSI, scale transformations are completely determined by the specification of $c$, $e$, and $f$.

\begin{figure}
\begin{center}
\plotone{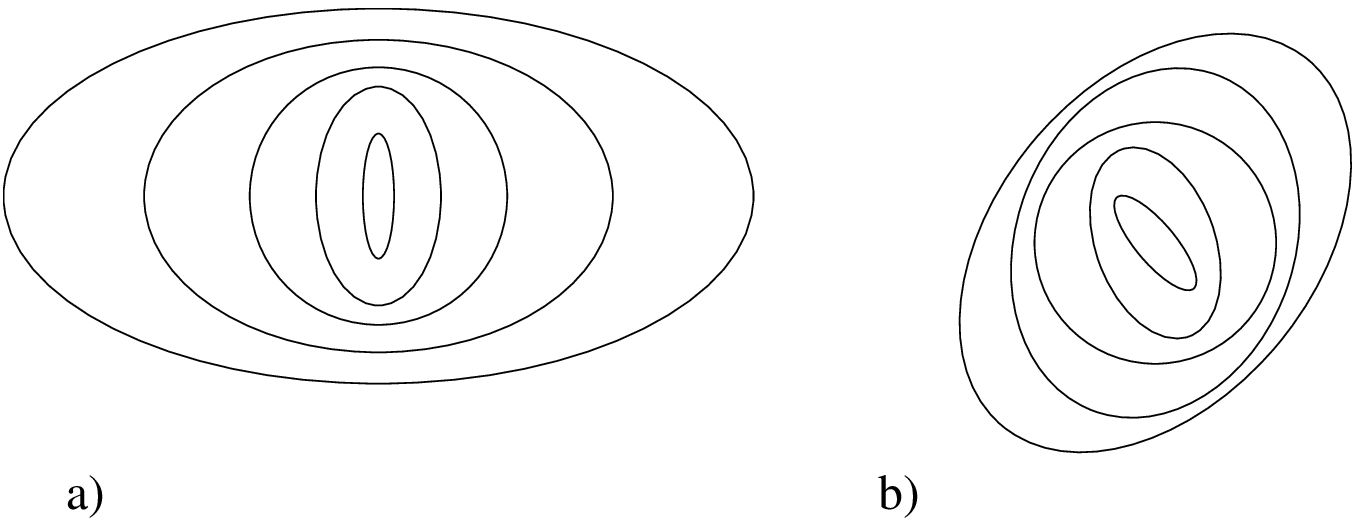}
\end{center}
\figcaption[f6.eps]{a) For self-affine scaling, one obtains ellipses that are stretched vertically for wave vectors of small magnitude, and horizontally elongated for large wave vectors. b) Ellipses are rotated in this example due to the presence of non-zero off-diagonal elements in the generator $G$.
\label{fig:balls}}
\end{figure}

It is instructive at this point to discuss a few examples of $G$, and corresponding family of balls obtained by repeated applications of $T_{\lambda}$ on the unit ball.
For the purpose of this discussion, we assume that $G$ is a 2x2 matrix, corresponding to the analysis of two-dimensional data with translationally invariant statistics. The simplest example is the case where $G$ is the identity matrix, corresponding to $T_{\lambda}=\lambda^{-1}$, or self-similar scaling. In addition, if the unit ball is chosen to be the unit circle, the family of balls generated by $T_{\lambda}$ are circles as well. 
More generally, the scaling may be different in two or more preferred directions. For simplicity, suppose there are two such directions assumed to coincide with the x and y axes, in which case the scaling is said to be self-affine: the off-diagonal elements of $G$ remain null, but the diagonal entries may be different from unity. If the unit ball is taken to be the unit circle, the corresponding balls are ellipses with principal axes pointing in the x and y directions. 
Furthermore, as illustrated in Figure~\ref{fig:balls}a), one finds that the balls are horizontally elongated for large wavenumbers ($k>1$) and vertically elongated for small ones ($k<1$). 
An example of physical systems presenting approximately such scaling are vertical cross-sections of the atmosphere where self-affinity is caused by the (stratifying) gravitational field of the Earth (Pflug et al. 1991, 1993). Finally, a matrix generator with non-zero off-diagonal elements indicates differential rotation in the scaling (see Figure~\ref{fig:balls}b)); in the atmosphere for instance, the observed differential rotation can be generated by the Coriolis force.

The numerical calculation of the anisotropy parameters $c$, $e$, and $f$ is performed using the Scale Invariant Generator (henceforth, SIG) technique (Lewis et al. 1999). Let us first define $P(\vec{k})$ to be the modulus squared of the Fourier transform of the field at wave vector $\vec{k}$. It follows from this definition that an angular integration of $P(\vec{k})$ yields $E(k)$, as defined in section~\ref{dtmtheory}. As a generalization of equation~\ref{eq:iso_scaling}, we have the scaling relation:

\begin{equation}
\langle P(\vec{k})\rangle \propto ||\vec{k}||^{-s}, \label{aniso_scaling}
\end{equation}
where the norm is with respect to $G$, the brackets indicate an average
over all realizations, and $s$ is a generalized scaling exponent given
in (isotropic) 2-D by $s=\beta+1$. It follows that $P(\vec{k})$ is in fact a function of $G$, $B_1$, $s$, and $\vec{k}$, that is, $P(\vec{k})\equiv P(c,e,f,B_1,s,\vec{k})$. If the analyzed sample consists of $N$ data points denoted by $\vec{k}_i$, $i=1,...,N$, we can define an error function $E_r$ by:

\begin{equation}
E_r^2(G) \equiv \frac{1}{N}\mathop{\sum}_{i,j}^{N}\left[ \ln P\left( \lambda_i^G\vec{k}_j \right) + s \ln \lambda_i - \ln P(\vec{k}_j) \right]^2, \label{eq:SIG}
\end{equation} 
\noindent where the sum is over the data points ($\vec{k}_j$) and scale ratio ($\lambda_i$). The anisotropic parameters are then estimated by minimizing this error function.

\section{Scale Invariant Generator Results} \label{sig}
The evaluation of the anisotropy exponents with the SIG technique
requires data over a wide range of scales (sufficient resolution), and good signal-to-noise ratio, such that the (approximate) isotropic scaling is valid over many scales.
As mentioned in section~\ref{data}, the images obtained at the OMM did not show sufficiently good statistics for the type of analysis described in this section.

\begin{figure}
\begin{center}
\plotone{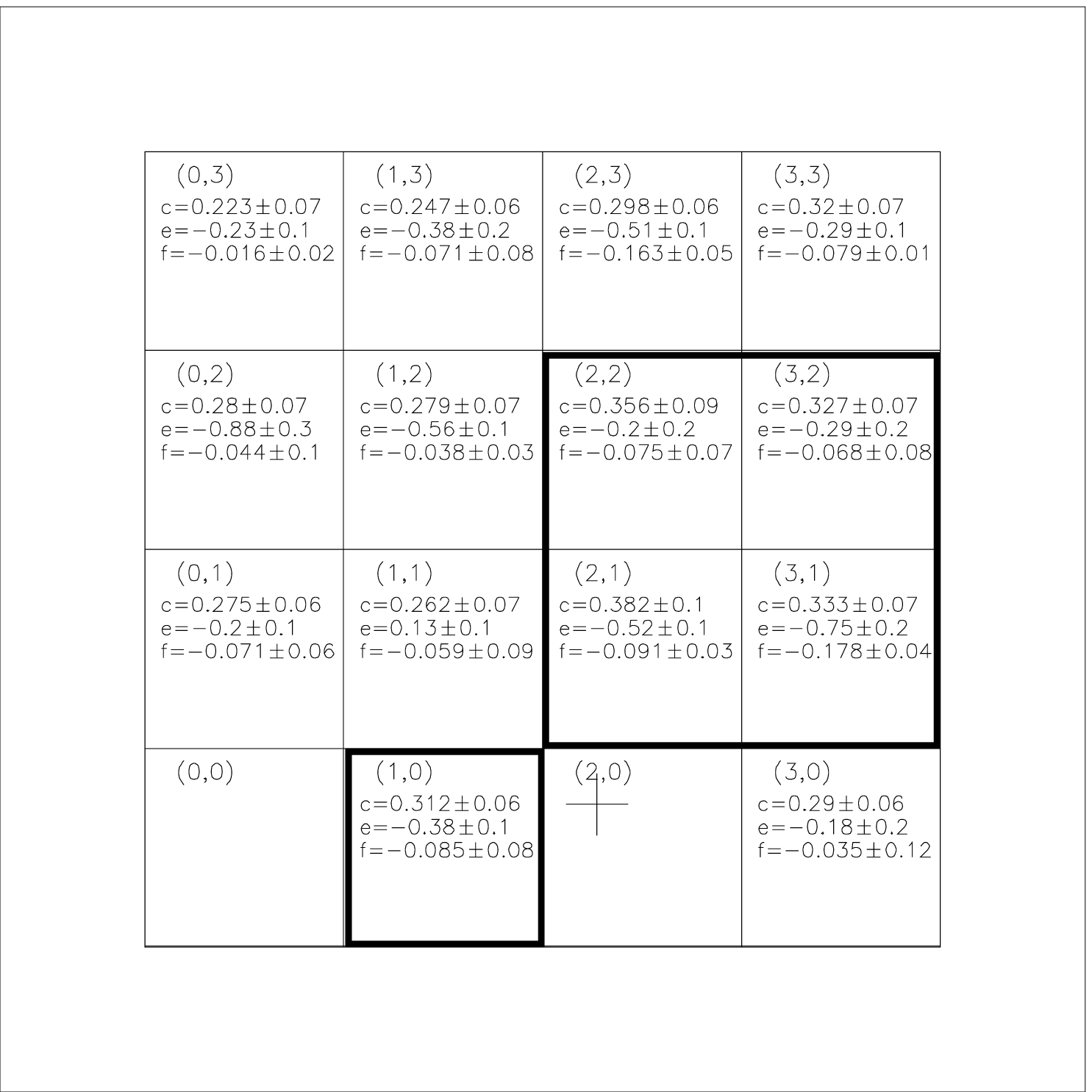}
\end{center}
\figcaption[f7.eps]{The results of the SIG analysis performed on the sub-images defined in Figure~\ref{fig:ggd18_image}. Boxes (1,0), (2,1), (3,1), (2,2) and (3,2) cover what is identified as the lobes, and are referred collectively as ``region I''. The remaining boxes belong to ``region II'', with the exception of (0,2) and (0,0) which contain the protostar and the western component of the binary GL 961, respectively.
\label{fig:gsiofggd18}}
\end{figure}

Figure~\ref{fig:gsiofggd18} displays the results of the SIG analysis performed on the sub-images of GGD 18 (defined in fig.~\ref{fig:ggd18_image}), where the quoted uncertainties on the parameters $c$, $e$, and $f$ were estimated from the numerical optimization of the error function $E_r$ defined in equation~\ref{eq:SIG}. 
Since the error surface near its minimum was typically wider in the $e$ direction, this exponent is known with less accuracy than the other two. Note that boxes (2,0) and (0,0) were left empty in Figure~\ref{fig:gsiofggd18} as they contain the source of GGD 18 and the western component of the binary GL 961. An immediate observation is the substantial variations of the parameters from one sub-region to the next; while the choice of linear GSI simplifies the numerical calculations, the fact that the generator $G$ is not constant over the region analyzed suggests that the framework of non-linear GSI may be more appropriate in this case.
We shall nevertheless assume that linear GSI is a good approximation over each subimage.

The parameters $e$ and $f$ which determine the off-diagonal elements of $G$ (eq.~\ref{eq:Gexpansion}), are non-zero for most sub-regions, hence providing evidence for the existence of differential rotation in the outflow dynamics. 
They are comparatively larger in sections (1,0), (2,1) and (3,1), which cover the portions of the southern and northern lobes that are the closest to the protostar--disk system.
As discussed in section~\ref{data}, boxes (1,2) and (2,3) contain parts of a jet of matter presumably emitted by GL 961 at some point in the past.
The large values of the off-diagonal elements of $G$ in these regions suggest that there is some rotation induced in the interaction region of this jet with GGD 18 material.
However, it is not clear at this moment where the rotation exactly occurs (e.g. in the jet, the perturbed material, or both) since the linear SIG technique doesn't resolve variations of the exponents within subimages.

It has been noted by Gravel (1990) that the eastern edge of the northern lobe of GGD 18 appears to be pushed westward presumably by the jet of material of GL 961. It is interesting to note that the region (1,1) where the jet and the northern lobe presumably comes the closest, is the only one with a positive value of $e$.

Let us divide the image into two subregions (see Figure~\ref{fig:gsiofggd18}) with region I covering what is morphologically identifiable as the lobes of the nebula,
and region II covering the rest of the observed field (with the exception of the emission features).
The assignation of subimages to either region was performed using the following criteria:  subregions belonging to region I (i) were close to the $30^o$ symmetry axis of the nebula, (ii) the corresponding light intensity was more important, and (iii) their intensity contours had conical shapes (indicative of ejection). 
With this classification, there appears to be a systematic difference in the mean value of $c$ between regions I and II: quantitatively, the difference in mean values, $\langle c_I \rangle - \langle c_{II}\rangle$ is found to lie 4.0 standard deviations away from the case $\langle c_I \rangle = \langle c_{II} \rangle$.
Although a precise identification of the stratifying factors is beyond the scope of our phenomenological framework, the observed stratification appears primarily dynamical in nature since $c$ is not observed to decrease monotonically with distance from the protostar (as  would be expected for gravitationally generated stratification), and since values of $c$ present clear differences within and outside the lobes of the nebula.

\section{Conclusion}

The statistics of scattered light in a small ensemble of YSOs have been shown to obey different scaling relations depending on the moment $q$ (``multiscaling''), with the relation between scaling laws and statistical moments given by $K(q)$. 
It was also shown that these $K(q)$ functions fit reasonably well into a
multifractal universality class, and that the corresponding universal
parameters are fairly uniform over the ensemble (with $\sigma_\alpha/\langle \alpha \rangle \sim 1\%$, $\sigma_{C_1}/\langle C_1 \rangle \sim 47\%$, and $\sigma_H/\langle H \rangle \sim 28\%$). 
Although universality may be thought of as describing the attractors of multiplicative cascades, and consequently is not sufficiently sensitive on the initial conditions of the cascade to allow the observation and characterization of fine details in the dynamics of a given multifractal field, the reasonable uniformity of the results presented 
suggests that the seven objects in the ensemble have similar (presumably turbulent) dynamics, as expected on theoretical grounds.

Finally, although obvious differences (e.g. the existence of a magnetic field) exist between the dynamics of the radiative fields of YSOs and that of terrestrial water clouds, our results suggest that the corresponding statistics are similar.
Indeed, the spectral slope and the universality parameters of atmospheric clouds ($\beta\approx 2.2, \alpha=1.79, H=0.63$ and $C_1=0.061$ -- Sachs et al. 1999) are close to that of bipolar nebulae (see table ~\ref{tab:results.all}).
Since empirically cloud liquid water behaves statistically approximately like a passive scalar (Lovejoy \&
Schertzer 1995), its statistical similarity with YSOs would make sense if dust grains in bipolar outflows constituted a passive scalar as well. Furthermore, the physics of radiative
transfer is also similar since in each case it is dominated by
scattering rather than absorption/emission processes.

All the objects in the ensemble presented reasonable isotropic scaling, that was 
systematically broken near the scale where noise becomes dominant in the 
measured signal. 
Only GGD 18 presented sufficient resolution to allow a GSI analysis of its statistics. 
Most of the sub-regions analyzed had matrix generators with non-zero off-diagonal elements, revealing the existence of differential rotation. 
The origin of the latter and its influence on mass ejection mechanisms should be accounted for in models of star formation; an obvious source of rotation in YSOs is the rotation of the disk-protostar system, and some of the properties of the coupling of this rotation to the ejected material might be studied by techniques similar to those used in this work.
All sub-images presented a non-zero value of $c$, indicating a possible stratifying force in the outflow mechanism. 
It was argued in section~\ref{sig} that such stratification would not be primarily gravitational in nature, but instead could result from dynamical pressure gradients related to physical forces, such as the centrifugal acceleration in the PP models, or the magnetic force in the Shu models. 
Finally, an important issue in the dynamics of GGD 18 is its interactions, if any, with the neighbouring YSO GL 961. Showing that such interactions are indeed involved would confirm that GGD 18 is located at approximately the same distance from the Earth as GL 961, namely 1.6 kpc. 
We found that the generator of (1,1) was the only one to have off-diagonal elements of opposite sign; as this box is believed to contain the region of closest approach between the northern lobe of GGD 18 and the jet from GL 961, the peculiarity of its anisotropy parameters could be a sign of interactions between the two YSOs.

It should be kept in mind that linear GSI is probably not accurate enough to probe fine details in the dynamics, and is in fact increasingly understood as measuring local multifractal textures (Pecknold et al. 1996, 1997). While developments in non-linear GSI or models involving non-scalar cascades, along with an increased ensemble of sufficient resolution, are probably necessary to obtain statistically robust statements concerning the outflow dynamics, it is hoped that our analysis has provided a foretaste of the vast possibilities of multiscaling analyses.

\acknowledgments
P.B. thanks the director of the Canada-France-Hawai telescope for a generous time allotment. S.L. was partly supported by the Natural Sciences and Engineering Research Council (NSERC) of Canada.

\begin{center}
REFERENCES
\end{center}
{\small
\noindent Appenzeller, I., Jankovics, I., \& Ostreicher, R. 1984, A\&A, {\bf 141}, 108
\\
\noindent Asselin, L., M\'enard, F., Bastien, P., Monin, J.-L., Rouan, D. 1996, ApJ, {\bf 472}, 349A
\\
\noindent Bachiller, R. 1996, ARA\&A, {\bf 34}, 111
\\
\noindent Bally, J., \& Lada, C. J. 1983, ApJ, {\bf 265}, 824
\\
\noindent Bastien, P., \& M\'enard, F. 1990, ApJ, {\bf 364}, 232
\\
\noindent Beckwith, S. V. W., Sargent, A. I., Koresko, C. D., \& Weintraub, D. A. 1989, ApJ, {\bf 343}, 393
\\
\noindent Biskamp, D. 1993, {\it Nonlinear Magnetohydrodynamics} (Cambridge: Cambridge University Press)
\\
\noindent Blitz, L. \& Thaddeus, P. 1980, ApJ, {\bf 241}, 676
\\
\noindent Bontemps, S., Andr\'e, P., Terebey, S., \& Cabrit, S. 1996, A\&A, {\bf 311}, 858
\\
\noindent Borgani, S., Murante, G., Provenzale, A., Valdarnini, R. 1993, Phys. Rev. E, {\bf 47}, 3879
\\
\noindent Cabrit, S. 1989, in {\it ESO Workshop on Low Mass Star Formation and Pre-Main Sequence Evolution}, ed. B. Reipurth (Garching: ESO), p. 119
\\
\noindent Cabrit, S., \& Andr\'e, P. 1991, ApJ, {\bf 379}, L25 
\\
\noindent Cadavid, A. C., Lawrence, J. K., Ruzmaikin, A. A., Kayleng-Knight, A. 1994, ApJ, {\bf 429}, 391
\\
\noindent Carbone, V., 1993, Phys. Rev. Lett., {\bf 71}, 1546
\\
\noindent --- \& Savaglio, S. 1996, MNRAS, {\bf 282}, 868
\\
\noindent ---, Veltri, P., Bruno, R. 1996, Nonlin. Proc. Geophys., {\bf 3}, 247 
\\
\noindent Cohen, M. 1973, ApJ, {\bf 185}, L75
\\
\noindent Coleman, P. H., \& Pietronero, L. 1992, Phys. Rep., {\bf 213}, 311
\\
\noindent Corrsin, S. 1951, J. Applied. Phys., {\bf 22}, 469
\\
\noindent Edwards, S., Cabrit, S., Strom, S. E., Heyer, I., Strom, K. M., \& Anderson, E. 1987, ApJ, {\bf 321}, 473  
\\
\noindent Ferreira, J. \& Pelletier, G. 1993a, A\&A, {\bf 276}, 625
\\
\noindent --- \& --- 1993b, A\&A, {\bf 276}, 637
\\
\noindent --- \& --- 1995, A\&A, {\bf 295}, 807
\\
\noindent Garrido, P., Lovejoy, S., Schertzer, D. 1996, Physics A, {\bf 225}, 294
\\
\noindent Gomez de Castro, \& A. I., Pudritz, R. E. 1992, ApJ, {\bf 395}, 501
\\
\noindent Gravel, P. 1990, M.Sc. thesis, Universit\'e de Montr\'eal, Montr\'eal, Qu\'ebec, Canada
\\
\noindent Gyulbudaghian, A. L., Glushkov, Y. I., Demisyuk, E. K. 1978, ApJ, {\bf 224}, L137
\\
\noindent Halsey, T. C., Jensen, M. H., Kadanoff. L. P., Procaccia, I., Schraiman, B. 1986, Phys. Rev. A, {\bf 33}, 1141
\\
\noindent Iroshnikov, P. 1963, Astron. Zh., {\bf 40}, 742
\\
\noindent Kraichnan, R. H. 1965, Phys. Fluids, {\bf 8}, 1385
\\
\noindent Lada, C. J. 1985, ARA\&A, {\bf 23}, 267
\\
\noindent ---, Gauthier III, T. N. 1982, ApJ, {\bf 261}, 161
\\
\noindent Lavall\'ee, D. 1991, Ph.D. thesis, McGill University, Montr\'eal, Canada
\\
\noindent ---, Schertzer, D., Lovejoy, S. 1991, in {\it Scaling, Fractals and Non-Linear Variability in Geophysics}, eds. D. Schertzer and S. Lovejoy (Klumer), p. 99
\\
\noindent ---, Lovejoy, S., Schertzer, D., Ladoy, P. 1993, in {\it Fractals in Geography}, eds. L. De Cola, N. Lam (Prentice Hall), p. 158
\\
\noindent Lenzen, R., Hodapp, K. W., Reddman, T. 1984, A\&A, {\bf 137}, 365 
\\
\noindent Lewis, G., Lovejoy, S., Schertzer, D., Pecknold, S. 1999, Computers in Geophys (in press)
\\
\noindent Longo, G., Vio, R., Paura, P., Provenzale, A., Rifatto, A. 1996, A\&A, {\bf 312}, 424
\\
\noindent Lovejoy, S., \& Schertzer, D. 1985, Wat. Resour. Res., {\bf 21}, 1233
\\
\noindent ---, --- 1990, Physics in Canada, {\bf 46}, 4, 46
\\ 
\noindent ---, Watson, B., Schertzer, D., Brosamlen, G. 1995, in {\it Particle Transport in Stochastic Media}, ed. L. Briggs (Portland: American Nuclear Society), p. 750
\\
\noindent ---, \& --- 1995, in {\it Fractals in Geoscience and Remote Sensing}, ed. G. Wilkinson (Luxembourg: Office for Official Publications of the European Communities), p. 102
\\
\noindent ---, Schertzer, D., Tessier, Y., 1998, Int. Journal. of Remote Sensing (submitted)
\\
\noindent ---, Garrido, P., Schertzer, D. 1998, Physica A (submitted).
\\
\noindent Mandelbrot, B. B. 1974, J. Fluid Mech., {\bf 62}, 331
\\
\noindent Martinez, V. J., Coles, P. 1994, ApJ, {\bf 437}, 550
\\
\noindent Naud, C., Schertzer, D., Lovejoy, S. 1996, in {\it Stochastic
Models in Geosystems}, eds. S. A. Molchansov and W. A. Woyczynski (Spinger-Verlag), p. 239
\\
\noindent Novikov, E. A., \& Stewart, R. 1964, Izv. Akad. Nauk. SSSR. Ser. Geofiz., {\bf 3}, 408
\\
\noindent Obukhov, A.  1949, Izv. Akad. Nauk. SSSR. Ser. Geogr. I Geofiz, {\bf 13}, 55
\\
\noindent Parisi, G., \& Frisch, U. 1985, in {\it Turbulence and Predictability in Geophysical Fluid Dynamics and Climate Dynamics}, eds. M. Ghil, R. Benzi, and G. Parisi (North-Holland), p. 72
\\
\noindent Pecknold, S., Lovejoy, S., Schertzer, D. 1996, in {\it
Stochastic Models in Geosystems}, eds. S. A. Molchansov and
W. A. Woyczynski (Spinger-Verlag), p. 269, 85
\\
\noindent Pecknold, S., Lovejoy, S., Schertzer, D., Hooge, C. 1997, in
{\it Scale in Geophysical Information Systems}, eds. D. Quattrochi and
M. F. Goodchild (Florida: CRC Press), p. 361
\\
\noindent Pelletier, G., \& Pudritz, R. E. 1992, ApJ, {\bf 394}, 117
\\
\noindent Pflug, K., Lovejoy, S., \& Schertzer, D. 1991, in {\it Nonlinear Dynamics of Structures}, eds. R. Z. Sagdeev, U. Frisch, A. S. Moiseev, and A. Erokhim (World Scientific), p. 72
\\
\noindent ---, ---, \& --- 1993, J.Atmos.Sci, {\bf 50}, 538
\\
\noindent Politano, H., \& Pouquet, A. 1995, Phys. Rev. E, {\bf 52}, 636
\\
\noindent Pompilio, M. P., Bouchet, F. R., Murante, G., Provenzale, A. 1995, ApJ, {\bf 449}, 1
\\
\noindent Richardson, L. F. 1922, {\it Weather prediction by numerical process} (Cambridge University Press)
\\
\noindent Sachs, Lovejoy, S., Schertzer, D. 1999 (in preparation for Fractals) 
\\
\noindent Schertzer, D., \& Lovejoy, S. 1983, in {\it Forth Symposium on Turbulent Shear Flows}, Karlshule, Germany
\\
\noindent ---, \& --- 1985, Phys. Chem. Hydrodyn., {\bf 6}, 623
\\
\noindent ---, \& --- 1987a, Ann. Sci. Math. du Qu\'ebec, {\bf 11}, 139
\\
\noindent ---, \& --- 1987b, J.Geophys.Res.D., {\bf 92}, 9693
\\
\noindent ---, \& --- 1989a, in {\it Fractals: Physical Origin and Consequences}, ed. L.Pietronero (Plenum), p. 49
\\
\noindent ---, \& --- 1989b, Pageoph, {\bf 130}, 57
\\
\noindent ---, \& --- 1991, in {\it Scaling, Fractals and Non-Linear Variability in Geophysics}, eds. D. Schertzer and S. Lovejoy (Kluwer), p. 41
\\
\noindent ---, \& --- 1995, in {\it Space/time Variability and Interdependence for various hydrological processes}, ed. R. A. Feddes (Cambridge: Cambridge University Press), p. 153
\\
\noindent ---, ---, Schmitt, F., Chigirinskaya, Y., Marsan, D. 1997, Fractals 5, 427
\\
\noindent ---, \& --- 1997a, J. Applied Meteorology, {\bf 36}, 1296
\\
\noindent ---, \& --- 1997b, ARM Proceedings 
\\
\noindent Schertzer, D., Schmitt, F., Naud, C., Marsan, D., Chigirinskaya, Y., Margeurite, C., Lovejoy, S., 1997, 7$\rm{th}$ Atmos. Rad. Meas. Sci. Team Meeting, 327-335
\\
\noindent Schmitt, F., Lavall\'ee, D., Schertzer, D., Lovejoy, S. 1992, Phys. Rev. Lett., {\bf 68}, 305
\\
\noindent Schmitt, F., Lovejoy, S., Schertzer, D. 1995, Geophys. Res. Lett., {\bf 22}, 1689
\\
\noindent Shu, F. H., Lizano, S., Ruden, S. P., Najita, J. 1988, ApJ, {\bf 328}, L19
\\
\noindent ---, Najita, J., Ostricker, E., Wilkin, F., Ruden, S., Lizano, S. 1994a, ApJ, {\bf 429}, 781
\\
\noindent ---, ---, Ruden, S., Lizano, S., 1994b, ApJ, {\bf 429}, 797
\\
\noindent Sylos Labini, F., Montuori, M. 1998, A\&A, {\bf 331}, 809
\\
\noindent Sylos Labini, F., Montuori, M., Pietronero, L., 1998, Phys. Rep., {\bf 293}, 726
\\
\noindent Turner, D. G. 1976, ApJ, {\bf 276}, 65
\\
\noindent Wiedenmann, G., Atmanspacher, H., Scheingraber, H. 1990, Can. Journ. Phys., {\bf 69}, 9, 827
\\
\noindent Yaglom, A. M. 1966, Sov. Phys. Dokl., {\bf 2}, 26
}
\end{document}